\documentstyle[aps,preprint,prl]{revtex}
\begin{document}  
\title{Intrinsic structure of two-phonon states in the
interacting boson model} 
\author{J.E.~Garc\'{\i}a--Ramos$^1$, C.E.~Alonso$^1$, 
J.M.~Arias$^1$, P.~Van Isacker$^2$ and  A.~Vitturi$^3$} 
\address{$^1$Departamento   de   F\'{\i}sica   At\'omica, Molecular  y
Nuclear, Universidad de Sevilla, Apartado 1065, 41080 Sevilla, Spain} 
\address{$^2$Grand Acc\'el\'erateur National d'Ions Lourds, B.P.~5027, 
F-14076 Caen Cedex 5, France} 
\address{$^3$Dipartimento di Fisica and INFN, Padova, Italy} 
\date{\today}
\maketitle
          
\begin{abstract}
A general  study of 
excitations up to two-phonon states is carried out
using   the intrinsic-state formalism of the Interacting Boson Model
(IBM). Spectra   and
transitions  for the different   dynamical symmetries are analyzed and
the correspondence  with states in the  laboratory frame is
established. The influence of multi-phonon states is discussed. The
approach is useful in
problems where  the complexity of the  IBM spectrum renders
the analysis in the laboratory frame difficult.
\end{abstract}

\pacs{PACS numbers: 21.60.-n, 21.60.Fw, 21.60.Ev}

\draft

\section{Introduction}

Whether double-$\gamma$ excitations exist or not in nuclei
has been under discussion during the last 30 years.
Double-$\gamma$ excitations 
correspond to $K^{\pi}=0^+$ and $K^{\pi}=4^+$ states,
and are the bandheads of two rotational bands
superimposed on the vibrational spectrum.
Recent experimental improvements in nuclear spectroscopy
following Coulomb excitation \cite{Fahl88},
inelastic neutron scattering \cite{Belg96}
and thermal-neutron capture \cite{Born93}
have made possible the study of highly excited low-spin states
which until now had been inaccessible.
In transitional even--even nuclei $^{186-192}$Os and $^{194}$Pt
large components of the double-$\gamma$ excitation
have been identified \cite{Wu83,Olln93}.
In well-deformed nuclei the search for double-$\gamma$ excitations
has led to the identification of a first candidate
namely the $K^{\pi}=4^+$ level at 2055 keV in $^{168}$Er
\cite{Born91}.
Subsequently, other two-phonon excitations have been observed.
For example, based on their decay properties the levels in $^{166}$Er
at excitation energies of 1943 and 2028 keV were judged to be collective
and interpreted as $K^{\pi}=0^+$ and $K^{\pi}=4^+$ members 
of the double-$\gamma$ multiplet, respectively \cite{Garr97,Fahl96}.
Other examples of double-$\gamma$ excitations
are the $K^\pi=4^+$ state at 1435 keV in $^{106}$Mo \cite{Gues95},
$K^\pi=4^+$ states in $^{154-156}$Gd \cite{Apra94} or 
$K^{\pi}=4^+$ state at 2173 keV in $^{164}Dy$ \cite{Corm97}.
It should be noted, however, that the identification of such states
as members of a two-phonon multiplet
is still very much under debate \cite{Burk94}.

A possible framework for investigating two-phonon excitations
is the Interacting Boson Model (IBM) \cite{Iach87}.
This model has been very successful
in describing low-lying collective spectra in medium-mass and heavy nuclei.
Two-phonon excitations are present in the IBM
but the analysis of their properties is not so easy
to carry out in the laboratory frame
since in the energy region of interest (around 2 MeV)
the density of states is high and their decay properties are intricate.
For example, it is as yet not clear how and, if so, to what extent
anharmonicities can be introduced in the model,
a question that did generate controversy
in the early days of the model \cite{Cast80,Bohr82}.

In this paper it is shown that
the intrinsic-state formalism \cite{Gino80,Diep80}
provides an ideal framework
for analysing the problem of two-phonon excitations in the IBM.
The intrinsic-state formalism
is an approximation (of order 1/N) to the exact IBM,
but its advantage is that it gives 
a clear interpretation of the structure of the states.
In order to address the problem of two-phonon excitations
we present here calculations in the Tamm--Dancoff Approximation (TDA),
extended up to two-particle--two-hole  excitations
carried in the intrinsic frame of the IBM.
This approximation is useful
due to its simplicity and its physical transparency,
and it allows to  derive analytical expressions
for energies and electromagnetic transitions
applicable to a wide range of situations.

The structure of the paper is as follows.
First, a brief recapitulation of the intrinsic-state formalism
as applied to the IBM is given,
which is subsequently extended to include up to two-phonon excitations.
Problems related to spurious states and anharmonicity are discussed.
In section~\ref{sch-cal} the results of schematic calculations
in the intrinsic frame are compared with those in the laboratory frame
for the three dynamical symmetry limits of the IBM. 
The character of the two-phonon states is established.
Section~\ref{el-trans} is devoted to electromagnetic transitions.  
Finally, in section~\ref{conclu} conclusions of this work are presented.

\section{The IBM in the intrinsic frame}
\label{ibmham}

The  IBM   describes low-lying collective
excitations in  even--even nuclei   in  terms  of bosons   with angular
momentum $0$ (s bosons) and $2$ (d  bosons) \cite{Iach87}.
These bosons interact via
a  hamiltonian that is rotational invariant and
number conserving
and usually includes  up to two-body interactions, 
although higher-order terms have been sometimes included.
The most  general 
two-body IBM hamiltonian may
be written in multipole form as 
\begin{equation}
\label{Ham1}
H= \varepsilon_s \hat {n}_s +  \varepsilon_d \hat {n}_d +
\kappa_0 \hat{P}^\dagger \,
\hat{P} + \sum_{L=1}^{4} \kappa_L \hat{T}_L\cdot\hat{T}_L,
\end{equation}
where
\begin{equation}
\label{P}
P^\dagger={1\over 2} d^\dagger\cdot d^\dagger -  
{1\over 2} s^\dagger \cdot s^\dagger
\end{equation}
and
\begin{equation}
\label{OpeMul}
\hat{T}_{L M}=
\sum_{\ell_1\ell_2}\;
\chi_{\ell_1\ell_2}^L
(\gamma^\dagger_{\ell_1}\times
\tilde{\gamma}_{\ell_2})_{LM},
\end{equation}
with
\begin{eqnarray}
\begin{array}{cccc}
\label{Chi}
\chi_{00}^1=0,&\chi_{02}^1=0,&\chi_{20}^1=0,&\chi_{22}^1=1\\
\chi_{00}^2=0,&\chi_{02}^2=1,&\chi_{20}^2=1,&\chi_{22}^2=\chi\\
\chi_{00}^3=0,&\chi_{02}^3=0,&\chi_{20}^3=0,&\chi_{22}^3=1\\
\chi_{00}^4=0,&\chi_{02}^4=0,&\chi_{20}^4=0,&\chi_{22}^4=1
\end{array}.
\end{eqnarray}
The symbol   $\cdot$ denotes scalar  product, in this work the scalar 
product is defined as $\hat T_L \cdot \hat T_L=
\sum\limits_M (-1)^{L-M} \hat T_{LM} \hat T_{L-M}$, while $\gamma$ with $\ell=0$
($\ell=2$) corresponds to an $s$ boson ($d$ boson).
In Eq.~(\ref{OpeMul}) the operator 
$\tilde{\gamma}_{\ell  m}=(-1)^{\ell-m}\gamma _{\ell-m}$
is introduced so as to verify the appropriate
properties with respect to spatial rotations. 

The  intrinsic-state   formalism  \cite{Gino80,Diep80}  provides a
connection  between  the IBM and  the Bohr--Mottelson collective model
\cite{Bohr75}.   Let us consider   a system consisting  of a
large  but  finite  number $N$ of   interacting bosons.  The dynamical
behaviour of this system can  be  described  in  lowest-order
as arising from independent bosons moving in an average field.
In this approximation,
the  ground state   of  such  a system is  a
condensate $|c\rangle$ of bosons,
all occupying a single state $\Gamma^\dagger_c$ of lowest energy,
\begin{equation}
\label{GS}
| c \rangle = {1 \over \sqrt{N!}} (\Gamma^\dagger_c)^N | 0 \rangle,
\end{equation}
where
\begin{equation}
\label{bc}
\Gamma^\dagger_c = {1 \over \sqrt{1+\beta^2}} \left (s^\dagger + \beta
\cos     \gamma          \,d^\dagger_0          +{1\over\sqrt{2}}\beta
\sin\gamma\,(d^\dagger_2+d^\dagger_{-2}) \right).
\end{equation}
The   dynamical variables   $\beta$ and  $\gamma$  are related  to the
quadrupole shape  variables  of  the geometrical  model  \cite{Bohr75}
($\beta\geq 0$). The expectation value of the IBM 
hamiltonian (\ref{Ham1}) in the ground state (\ref{GS}) is
\begin{eqnarray}
\label{Ener1}
\langle c|H|c\rangle&=&
{\displaystyle {N\over{5(1+\beta^2 )}}}
\Bigl(5\,\varepsilon_s+25\,\kappa_2
+\beta^2\,(5\,\varepsilon_d-3\,\kappa_1 
+5\,\kappa_2+5\,\chi^2\,\kappa_2-7\,\kappa_3+9\,\kappa_4)\Bigr)
\nonumber\\
&+&{\displaystyle {\frac{N(N-1)}{140{{(1+\beta^2)}^2}}}}
\Bigl(35\,\kappa_0+\beta^2(-70\,\kappa_0+560\,\kappa_2)
-80\,{\sqrt{14}}\,\beta^3\,\chi\,\cos(3\,\gamma)\,\kappa_2
\nonumber\\
\qquad\qquad\qquad
&&+\beta^4(35\,\kappa_0+40\,\chi^2\,\kappa_2+72\,\kappa_4)\Bigr).
\end{eqnarray}
Alternative expressions can be found in  Refs.~\cite{Gino80,Isac81,Levi87}
using  normal-ordered hamiltonians.
The  equilibrium values of the  deformation parameters are obtained by
minimizing  the energy expression (\ref{Ener1})  with   respect to
$\beta$ and  $\gamma$.   Due to physical   considerations  a
minimum  for  $\beta\rightarrow\infty$    must  be excluded
\cite{Levi87}.
Furthermore it follows from (\ref{Ener1})
that a  triaxial  minimum is excluded
and only prolate ($\gamma=0^o$) or oblate ($\gamma=60^o$) solutions
are allowed.
For an attractive quadrupole force ($\kappa_2<0$)
these are found for $\chi<0$ and $\chi>0$, respectively.
A special situation of 
$\gamma$-unstability occurs if the quadrupole term vanishes ($\kappa_2=0$)
or if $\chi=0$.

The condensed boson $\Gamma_c^\dagger$ (\ref{bc}) is the first 
component of a new boson basis (deformed or intrinsic bosons). 
These deformed bosons
are related to the spherical ones (laboratory bosons) through a
general  unitary transformation $\eta$,
\begin{equation}
\label{Har1}
\Gamma    _p^\dagger=   \sum_{\ell     m}\eta_{\ell   m}^p\gamma_{\ell
m}^\dagger,    \qquad    \gamma_{\ell   m}^\dagger=   \sum_p\eta_{\ell
m}^{p*}\Gamma_{p}^\dagger,
\end{equation}
the  deformation   parameters  $\eta_{\ell  m}^{p}$
verifying the orthonormalization conditions
\begin{equation}
\label{orto}
\sum_{\ell  m}  \eta_{\ell  m}^{p'*}  \eta_{\ell m}^{p} =\delta_{pp'},
\qquad    \sum_{p}      \eta_{\ell      m}^{p*}     \eta_{\ell'm'}^{p}
=\delta_{\ell\ell'} \delta_{mm'}.
\end{equation}
The index $p$ labels the different deformed bosons.
After obtaining the $\eta$  parameters  for $p=c$
({\it i.e.}~the equilibrium deformation  parameters),
the $\eta$ parameters for the excited
bosons follow from the   orthogonality   conditions
(\ref{orto}).  An  appropriate  choice of   deformed bosons can be
found  in  Ref.~\cite{Levi87}:
\begin{eqnarray}
\begin{array}{c}
\label{etas}
\eta_{00}^c={\displaystyle{1\over\sqrt{1+\beta^2}}}~~,~~
\eta_{20}^c={\displaystyle
{\beta\cos\gamma\over\sqrt{1+\beta^2}}}~~,~~
\eta_{22}^c=\eta_{2-2}^c={\displaystyle
{\beta\sin\gamma\over\sqrt{2}\sqrt{1+\beta^2}}}                    \\
\eta_{00}^\beta={\displaystyle {-\beta\over\sqrt{1+\beta^2}}}~~,~~
\eta_{20}^\beta={\displaystyle
{\cos\gamma\over\sqrt{1+\beta^2}}}~~,~~
\eta_{22}^\beta=\eta_{2-2}^\beta={\displaystyle
{\sin\gamma\over\sqrt{2}\sqrt{1+\beta^2}}}                          \\
\eta_{20}^{\gamma+}={\displaystyle
{-\sin\gamma}}~~,~~
\eta_{22}^{\gamma+}=
\eta_{2-2}^{\gamma+}={\displaystyle{\cos\gamma\over\sqrt{2}}}  \\
\eta_{21}^{x}={\displaystyle{1\over\sqrt{2}}}~~,~~
\eta_{2-1}^{x}={\displaystyle{1\over\sqrt{2}}}          \\
\eta_{21}^{y}={\displaystyle{1\over\sqrt{2}}}~~,~~
\eta_{2-1}^{y}={\displaystyle{-1\over\sqrt{2}}}            \\
\eta_{22}^{\gamma-}={\displaystyle{1\over\sqrt{2}}}~~,~~
\eta_{2-2}^{\gamma-}={\displaystyle {-1\over\sqrt{2}}}
\end{array},
\end{eqnarray}
other coefficients being zero.  This basis  has a clear
correspondence with the usual $\beta$ and $\gamma$ excitations.

In the following section the usefulness of this notation for 
the calculation of matrix elements in a one- and two-phonon
basis shall become clear.

\subsection{One- and two-phonon states}

The excited bands can be considered as intrinsic excitations
built on the boson condensate (\ref{GS}), obtained by replacing a 
$c$ boson by an excited one, two $c$ bosons by two excited ones and so on.
This set of excitations contains spurious states
that are coupled to the true physical ones
and must be removed \cite{Levi87}.
This problem is a consequence of TDA.
In the Random Phase Approximation (RPA)
all the spurious states come out with zero energy
and are decoupled from the physical states \cite{Duke84}.
However, RPA is more difficult to implement than TDA and
obscures any physical interpretation. The problem of the 
spurious states will be returned to later.

The excited states included in the present analysis are
\begin{eqnarray}
\begin{array}{l}
\label{exst}
| p \rangle = {\displaystyle{1\over\sqrt{(N-1)!}}}\Gamma^\dagger_p 
({\Gamma^\dagger}_c)^{N-1}| 0 \rangle,\\
| p\,p' \rangle ={\displaystyle{1 \over \sqrt{1 + \delta_{p,p'}}}}
{\displaystyle{1\over\sqrt{(N-2)!}}} {\Gamma^\dagger}_p 
{\Gamma^\dagger}_{p'} ({\Gamma^\dagger}_c)^{N-2}|0\rangle, 
\end{array}
\end{eqnarray}
where $p,p'\neq c$. Eigenfunctions are linear 
combinations of these excited states plus the ground state,
\begin{equation}
\label{eigfun}
| \phi^\xi \rangle = N^\xi \left( W^\xi |c\rangle + 
\sum\limits_{p} {X^\xi}_p |p\rangle +
\sum\limits_{p p'} {Y^\xi}_{p p'}|pp'\rangle \right),
\end{equation}
where $N^\xi$ is a normalization constant. 
The ground-state contribution comes from the off-diagonal matrix elements of
the hamiltonian between the ground state and the two-phonon states.
  
To obtain the states (\ref{eigfun}),
the hamiltonian is diagonalized in the basis of excited 
states (\ref{exst}) plus the ground state (\ref{GS}). 
The different matrix elements that intervene are
\begin{equation}
\label{f1}
\langle c | H | p \rangle = \sqrt{N} F^{(1)}_{cp} +
                           2 \sqrt{N} (N-1) F^{(2)}_{cccp}=0,
\end{equation}

\begin{equation}
\label{f2}
\langle c |H | p p'\rangle =
{\displaystyle{2 \sqrt{N}\sqrt{N-1}\over\sqrt{1+\delta_{pp'}}}} 
F^{(2)}_{ccpp'},
\end{equation}

\begin{equation}
\label{f3}
\langle p |H | p' \rangle =F^{(1)}_{pp'}+4(N-1)F^{(2)}_{cpp'c}+ 
            \delta_{pp'}(N-1) \left(F^{(1)}_{cc}+(N-2)F^{(2)}_{cccc}\right), 
\end{equation}

\begin{eqnarray}
\label{f4}
\nonumber
\langle p |H | p'p'' \rangle =
{\displaystyle{\sqrt{N-1}\over\sqrt{1+\delta_{p'p''}}}}   
\left(4 F^{(2)}_{cpp'p''} \right.
& + & \left. \delta_{pp'}F^{(1)}_{cp''}+ \delta_{pp''}F^{(1)}_{cp'} \right.\\
& + & \left. 2(N-2)(\delta_{pp'}F^{(2)}_{cccp''}+\delta_{pp''}F^{(2)}_{cccp'})
\right),
\end{eqnarray}

\begin{eqnarray}
\label{f6}
\nonumber
\langle p p' |H | p''p''' \rangle & = &
{\displaystyle {\delta_{pp''}\delta_{p'p'''}+\delta_{pp'''}\delta_{p'p''} 
\over 1 + \delta_{pp'} \delta_{pp''} \delta_{p''p'''}}}
\left((N-2)F^{(1)}_{cc}+(N-2)(N-3)F^{(2)}_{cccc}\right) \\
\nonumber
& + & {\displaystyle {1 \over \sqrt{(1 + \delta_{pp'}) (1 + \delta_{p''p'''})}}} 
\left\{ 4 F^{(2)}_{pp'p''p'''}
+\delta_{pp''}\left(F^{(1)}_{p'p'''}+4(N-2)F^{(2)}_{cp'p'''c}\right) \right. \\
\nonumber
& + & \left.\delta_{pp'''}\left(F^{(1)}_{p'p''}+4(N-2)F^{(2)}_{cp'p''c}\right)
+\delta_{p'p''}\left(F^{(1)}_{pp'''}+4(N-2)F^{(2)}_{cpp'''c}\right) \right. \\
& + & \left.\delta_{p'p'''}\left(F^{(1)}_{pp''}+4(N-2)F^{(2)}_{cpp''c}\right)
\right\}~~,
\end{eqnarray}
where
\begin{equation}
\label{F1}
F^{(1)}_{pp'}=\sum\limits_{\ell_1 m_1} \tilde\varepsilon_{\ell_1} 
              \eta_{\ell_1 m_1}^{p*}\eta_{\ell_1 m_1}^{p'}
\end{equation}
and
\begin{equation}
\label{F2}
F^{(2)}_{p_1 p_2 p_3 p_4}=
\sum\limits_{\ell_1 m_1 \ell_2 m_2 \ell_3 m_3 \ell_4 m_4} 
V_{\ell_1 m_1, \ell_2 m_2, \ell_3 m_3, \ell_4 m_4}
\eta_{\ell_1 m_1}^{p_1*}\eta_{\ell_2 m_2}^{p_2*}
\eta_{\ell_3 m_3}^{p_3}\eta_{\ell_4 m_4}^{p_4}.
\end{equation}
The coefficients $\tilde\varepsilon_{\ell}$ include the single-particle
energies $\varepsilon_{\ell}$ plus 
contributions from the two-body terms
in the IBM hamiltonian (\ref{Ham1}). The coefficients
$V_{\ell_1 m_1, \ell_2 m_2, \ell_3 m_3, \ell_4 m_4}$
are defined as 
\begin{equation}
V_{\ell_1 m_1, \ell_2 m_2, \ell_3 m_3, \ell_4 m_4} \equiv
{1\over4}\langle \ell_1 m_1, \ell_2 m_2 | V |\ell_3 m_3, \ell_4 m_4 \rangle
\sqrt{1+\delta_{\ell_1 \ell_2}\delta_{m_1 m_2}}
\sqrt{1+\delta_{\ell_3 \ell_4}\delta_{m_3 m_4}}~.
\end{equation}
By construction $F^{(1)}_{pp'}$ is symmetric under interchange of $p$ and 
$p'$  and $F^{(2)}_{p_1 p_2 p_3 p_4}$ has the same symmetry properties 
as $V_{\ell_1 m_1, \ell_2 m_2, \ell_3 m_3, \ell_4 m_4}$.

\subsection{Spurious modes. Goldstone bosons}

Before diagonalization spurious bosons must be removed from the basis. 
The procedure to separate out those modes from 
the physical ones is described in detail in Ref.~\cite{Levi87}.
In this section we just enumerate the spurious (Goldstone)
modes that occur in the different geometrical limits of the IBM. 

\begin{itemize}
\item $\beta=0$, spherical nuclei. 
\begin{itemize}
\item $O(3)$ or $O(5)$ scalar hamiltonian. There are no Goldstone bosons.
\item $O(6)$ scalar hamiltonian. A spherical nucleus corresponds to 
the $U(5)$ dynamical symmetry, not compatible with 
a symmetry group $O(6)$. Therefore this possibility must be excluded.
\end{itemize}
\item $\beta>0$, deformed nuclei.
\begin{itemize}
\item $O(3)$ scalar hamiltonian. For $\gamma=0$, $\Gamma_x$ and $\Gamma_y$ are
spurious bosons. For $\gamma=\pi/3$ the spurious bosons are $\Gamma_x$ and
$\Gamma_{\gamma-}$. In the triaxial case
$\Gamma_x$, $\Gamma_y$ and $\Gamma_{\gamma-}$ are Goldstone bosons.
\item $O(5)$ scalar hamiltonian. All the excited bosons are spurious
except the $\Gamma_\beta$ boson.
\item $O(6)$ scalar hamiltonian. All the excited bosons correspond to
Goldstone bosons.
\end{itemize}
\end{itemize}

\subsection{The anharmonicity problem}
\label{anh-prob}

It is not clear {\it a priori} to what extent $\beta$ and $\gamma$
vibrations are harmonic in the IBM.
The analysis of this problem is quite difficult in the laboratory frame;
with the intrinsic-state formalism, however, some general conclusions
can be obtained for arbitrary hamiltonians.

Let us consider a system with a large number of bosons $N$. For large $N$
the IBM hamiltonian is approximately diagonal in the basis of
excited states (\ref{exst}). The degree of anharmonicity $\varsigma$ can be
defined as (with this definition the harmonic limit corresponds to 
$\varsigma=0$)
\begin{equation}
\label{anhar}
\varsigma = {E^{ex}_{p^2} \over E^{ex}_p} - 2 ,
\end{equation} 
with 
\begin{equation} 
\label{ex-p}
E^{ex}_{p}=\langle p | H |p \rangle-\langle c | H | c\rangle,
\end{equation}
\begin{equation}
\label{ex-pp} 
E^{ex}_{p^2}=\langle p p | H |p p\rangle-\langle c | H | c\rangle ~.
\end{equation} 
These can be evaluated with the help of the following expressions, in which
a Hamiltonian including up to three--body interactions have been supposed, 
\begin{equation}
\label{ener-gs3}
\langle c | H | c\rangle=  N(N-1)(N-2) F^{(3)}_{cccccc} 
                         + N(N-1) F^{(2)}_{cccc}+ N F^{(1)}_{cc},
\end{equation} 

\begin{eqnarray}
\label{ener-p3}
\nonumber
\langle p | H | p\rangle=(N-1)(N-2)(N-3)F^{(3)}_{cccccc} 
& + & (N-1)(N-2) F^{(2)}_{cccc}+ 9(N-2)(N-3)F^{(3)}_{ccppcc} \\
& + & (N-1) \left(F^{(1)}_{cc} +4 F^{(2)}_{pccp}\right) +F^{(1)}_{pp},
\end{eqnarray} 

\begin{eqnarray}
\label{ener-pp3}
\nonumber
\langle p p | H | p p \rangle= (N-2)(N & - & 3)(N-4)F^{(3)}_{cccccc}
+ (N-2)(N-3) \left(F^{(2)}_{cccc}+18F^{(3)}_{ccppcc}\right) \\
& + & (N-2) \left(F^{(1)}_{cc} + 8 F^{(2)}_{pccp}+18 F^{(3)}_{cppppc} \right)
+ 2 F^{(1)}_{pp}+ 2 F^{(2)}_{pppp},
\end{eqnarray} 
where $F^{(1)}$ and $F^{(2)}$ are defined in Eqs.~(\ref{F1}) and 
(\ref{F2}), and  $F^{(3)}$ equals
\begin{equation}
\label{F3}
F^{(3)}_{p_1 p_2 p_3 p_4 p_5 p_6}=
\sum\limits_{\ell's, m's}
U_{\ell_1 m_1, \ell_2 m_2, \ell_3 m_3, \ell_4 m_4, \ell_5 m_5, \ell_6 m_6}
~\eta_{\ell_1 m_1}^{p_1*}\eta_{\ell_2 m_2}^{p_2*}
\eta_{\ell_3 m_3}^{p_3*}\eta_{\ell_4 m_4}^{p_4}
\eta_{\ell_5 m_5}^{p_5}\eta_{\ell_6 m_6}^{p_6},
\end{equation}
where $U_{\ell_1 m_1, \ell_2 m_2, \ell_3 m_3, \ell_4 m_4, \ell_5 m_5, 
\ell_6 m_6}$ are the interaction matrix elements
between symmetrized, normalized three-boson states.

With the expressions (\ref{anhar}--\ref{ener-pp3})
it can be shown that the leading order of $\varsigma$ is $1/N$. 
Furthermore, this result remains valid
for a general IBM hamiltonian with up to $n$-body interactions.
The conclusion is thus that anharmonicities
can only exist in the IBM for a finite number of bosons.
Only in the U(5) limit \cite{Arim76},
when some of the coefficients (\ref{ener-gs3}--\ref{ener-pp3}) exactly vanish,
the spectrum can be anharmonic even for infinite $N$.
The harmonicity of the IBM spectra in the limit $N\rightarrow\infty$
will be used as a guide
to identify the number of excited bosons in a given state.

\section{Schematic calculations}
\label{sch-cal}
\subsection{Dynamic symmetries}

The analysis of the  three dynamical symmetry limits of the IBM  
provides a good test of the formalism presented in the previous section.

In  calculations carried  out in  the laboratory frame
it is, in  some cases, difficult to identify the  character of a band
($\beta$, $\gamma$, $\dots$).
This becomes increasingly difficult as the number of phonons increases.
In the intrinsic-frame calculations, on the other hand,
the character of the band is easily identified
from the structure of the state.
For high $N$ the intrinsic-state formalism
provides a good approximation to the exact calculation
and it can thus be used to clearly identify the character of the bands
in the laboratory frame.
For low $N$ the character of the bands in the laboratory frame
can be obtained extrapolating the results of high $N$.

In the following
the intrinsic-state results for the three dynamical 
symmetry limits of IBM are presented.

\begin{itemize}
\item
$U(5)$ limit. The hamiltonian in this limit is
\begin{equation}
\label{hamu5}
H= \varepsilon_d \hat {n}_d + \sum_{L=1,3,4} \kappa_L
\hat{T}_L \cdot \hat{T}_L .
\end{equation} 
It corresponds  to the spherical limit ($\beta=0$).
There are no Goldstone bosons.
The ground state is a condensate of $s$ bosons $(s^\dagger)^N |0\rangle$
and coincides with  the exact (laboratory) ground state.
Excited bosons correspond to $d^\dagger$;
therefore excited states are equal in the intrinsic and laboratory frames.
A calculation in the intrinsic frame thus coincides with the exact one.
 
\item
$SU(3)$ limit.
This dynamical symmetry corresponds to the hamiltonian
\begin{equation} 
\label{hamsu3}
H= \sum_{L=1,2}\kappa_L \hat{T}_L\cdot\hat{T}_L.
\end{equation} 
with $\chi=\pm\sqrt{7}/2$.
For prolate nuclei $\Gamma_{x}$ and $\Gamma_{y}$
are the only Goldstone bosons
while, for oblate nuclei,
these are $\Gamma_{x}$ and $\Gamma_{\gamma-}$.
The analysis of the character of the bands
is especially relevant in this case
because the concepts of $\gamma$ and $\beta$ bands
are appropriate for this limit
or situations close to it, {\it i.e.}~well-deformed nuclei. 

The character of the bands built on one- and two-phonon excitations
traditionally has been identified in the laboratory frame
in terms of the quantum number $K$
and the most symmetric representations of $SU(3)$ \cite{Arim78}.
In this way the $\beta$ band
corresponds to the representation $(2N-4,2)~K=0$
[where $(\lambda,\mu)$ are the Elliott quantum numbers]
and the $\gamma$ band to $(2N-4,2)~K=2$.
In the case of two-phonon excitations
assignments are already less clear.
There are four possible two-phonon bands:
two with $K=0$ ($\beta^2$ and $\gamma^2$),
one with $K=2$ ($\beta\gamma$)
and one with $K=4$ ($\gamma^2$).
These correspond to the $(2N-8,4)$ and the $(2N-6,0)$
representations of the $SU(3)$ limit.
It is not {\it a priori} clear, however,
how the bands should be assigned to the $SU(3)$ representations;
in particular, this ambiguity exists for the $K=0$ bands.
We now proceed to show how this problem can be solved
using the intrinsic-state formalism.

It is worth noting that with the definition of $\gamma+$ and $\gamma-$
bosons given in Eq.~(\ref{etas}) and the definition of two--phonon states
given in Eq.~(\ref{exst}), the states with two $\gamma+$ or two $\gamma-$
excitations have no good $K$ quantum number for the case of axial symmetry. 
In order to have good $K$ in that case, a simple linear combination 
correctly normalized can be done,

\begin{equation}
\label{g2K0}
|\gamma^2_{K=0} \rangle = {1 \over \sqrt{2}}\left[|\gamma_+^2 \rangle
+ |\gamma_-^2 \rangle \right] ~,
\end{equation}

\begin{equation}
\label{g2K4}
|\gamma^2_{K=4} \rangle = {1 \over \sqrt{2}}\left[|\gamma_+^2 \rangle
- |\gamma_-^2 \rangle \right] ~. 
\end{equation}
 
We will work with these states even in the cases in which $K$ is considered 
not to be a good quantum number.

In Fig.~\ref{fig-su3}
we present a calculation for a particular $SU(3)$ hamiltonian
and compare the exact and approximate calculations for a wide range of $N$.
In general, to identify the character of an excitation
its energy in the intrinsic and laboratory frame should be compared.
This procedure is not unambiguous for low $N$.
Therefore, first a large $N$ is considered
in which case the excitation energies
are proportional to the number of excited phonons,
{\it i.e.}~the spectrum is harmonic (see section~\ref{anh-prob}).
For $\kappa_1=0$
the eigenenergies of the hamiltonian (\ref{hamsu3}) are
\begin{equation}
\label{enersu3}
E(L,\lambda,\mu)=-{3\kappa_2\over8}L(L+1)+
{\kappa_2\over2}(\lambda^2+\mu^2+\lambda\mu+3\lambda+3\mu).
\end{equation}
The $(\lambda,\mu)$ contained in a symmetric representation $[N]$ of $U(6)$
are
\begin{equation}
\label{rep-su3}
(\lambda,\mu)=(2N-4n_x-6n_y,2n_x),\qquad n_x,n_y=0,1,2,...,
\end{equation}  
with $2N-4n_x-6n_y\geq0$.
Neglecting the $L(L+1)$ term in (\ref{enersu3}),
the excitation energies for the different representations are 
\begin{equation} 
\label{ex-su3}
E(\lambda,\mu)-E(2N,0)=-\kappa_2 (6 n_x+12n_y) N +\kappa_2 (6 n_x^2
+18 n_y^2+18 n_x n_y-3 n_x-9n_y).
\end{equation} 
From this expression it follows
that $\Delta n_x=+1$ corresponds to a single-phonon excitation
and $\Delta  n_y=+1$ to a double-phonon excitation.
So it is clear that the one-phonon excitations
are in the $(2N-4,2)$ representation
while the two-phonon excitations belong to $(2N-8,4)$ or $(2N-6,0)$. 

Finally, to fully determine the character of the states
approximate and exact energies are compared
and the structure of the state is determined in the intrinsic frame.
In this way the mixture
between the $\beta^2$ and $\gamma_{K=0}^2$ excitations is found
and the result is
that $(2N-8,4)K=0$ corresponds to the combination
$\sqrt{2\over3}\beta^2+\sqrt{1\over3}\gamma_{K=0}^2$
while $(2N-6,0)K=0$ corresponds to
$\sqrt{1\over3}\beta^2-\sqrt{2\over3}\gamma_{K=0}^2$.
If the quadrupole generator of $SU(3)$ is used as E2 operator,
no transitions can occur between the two bands
because they belong to different $SU(3)$ representations.
This result is reproduced in the intrinsic frame
since their particular composition prevents E2 transitions between them
(see section~\ref{el-trans}).

\item
$O(6)$ limit. The hamiltonian for this limit is
\begin{equation}
\label{hamo6}
H= \kappa_0 \hat{P}^\dagger \,
\hat{P} + \sum_{L=1,3} \kappa_L \hat{T}_L\cdot\hat{T}_L.
\end{equation}
A particular case of this hamiltonian is
\begin{equation} 
H=\kappa_2' \hat{T}_2\cdot\hat{T}_2,
\end{equation} 
with $\chi=0$.
This latter form is particularly useful
in the study of the transition from $SU(3)$ to $O(6)$
by varying a single parameter $\chi$.

In this limit the concept of $\beta$ and $\gamma$ bands
is quite vague
and it is therefore difficult to find the bandheads in the spectrum
and to identify their character.
To determine the number of excited phonons
in a given state of the $O(6)$ limit,
the harmonicity of the spectrum for high boson number is used
(see section~\ref{anh-prob}).
States are labeled by $(\sigma,\nu_\Delta,\tau,L)$ \cite{Arim79}
and the possible values of $\sigma$ and $\tau$ are,
\begin{eqnarray}
\sigma&=&N-2n>0,\qquad n=0,1,...,
\nonumber\\
\tau&=&\sigma,\sigma-1,...,1,0.
\end{eqnarray}
The eigenenergies in terms of these labels are
\begin{equation}
\label{o6-ener}
E(\sigma,\tau,L)=A \sigma (\sigma+4)+B \tau (\tau+3) + C L(L+1),
\end{equation} 
with for realistic calculations $A<0$, $B>0$ and $C>0$.
For these parameters, bandheads correspond to $\tau=0$ and $L=0$.
Their energy is given by
\begin{equation}
E(\sigma)= A \sigma (\sigma+4)
\end{equation}       
and their excitation energy is,
\begin{equation}
E(\sigma=N-2n)-E(\sigma=N)= 4A (- n N ) + 4A(n^2-2n). 
\end{equation} 
From the latter expression it follows
that $n$ represents the number of excited phonons.

In Fig.~\ref{fig-o6}
a calculation for a particular $O(6)$ hamiltonian is shown
and a comparison of the exact and approximate results
for a wide range of $N$ is presented. 
This limit corresponds to the $\gamma$-unstable vibrator
and the only physical excited boson is $\Gamma_\beta$.
\end{itemize}

\subsection{Transitional hamiltonians}

In the previous section
the degree of accuracy of the method was illustrated.
In this section schematic hamiltonians are analyzed
which do not correspond to any of the three dynamical symmetry limits.
A fixed number of bosons is used
and results are plotted as a function of a parameter in the hamiltonian
that allows to explore a wide range of situations,
including presumably unrealistic ones.
In these calculations the Goldstone bosons
correspond to $\Gamma_x$ and $\Gamma_y$
and have been removed before diagonalization.
For some extreme values of the parameters,
the ground state contains a sizeable contribution from two-phonon excitations.

Figure~\ref{ppqq}a shows the results of a calculation
with a quadrupole and a pairing interaction 
($H=\kappa_0 \hat P^\dagger \hat P + \kappa_2 \hat T_2 \cdot \hat T_2$, 
with $\chi=-\sqrt{7}/2$) for $N=16$.
It is seen that the character of the different bands changes
as a function of the ratio $-\kappa_0/\kappa_2$.
Note the increase with $-\kappa_0/\kappa_2$ of the energy of the $\beta$ band
while the $\gamma$-band energy remains approximately constant
and the harmonic behaviour between the one-$\gamma$ and two-$\gamma$ bands
holds throughout.
This figure displays the same trends as Fig.~14a of Ref. \cite{Cast88}
which shows an identical calculation in the laboratory frame.
Only small differences occur
due to mixing with higher-order phonon excitations.
Note that a realistic value of the ratio $-\kappa_0/\kappa_2$ is around $4$.

Figure~\ref{t4t4qq}a shows the influence of a hexadecapole term added to
a quadrupole term ($H=\kappa_2 \hat T_2 \cdot \hat T_2
+ \kappa_4 \hat T_4 \cdot \hat T_4$, with $\chi=-\sqrt{7}/2$)
for $N=16$. In the whole range of variation
of the parameter $\kappa_4/\kappa_2$ there is a clear separation
in energy and only small mixing between the
one- and two-phonon states.
The $SU(3)$ limit is recovered for $\kappa_4=0$.

In Fig.~\ref{edqq}a a transition from a deformed to a spherical
nucleus is analyzed via the hamiltonian $H=\varepsilon_d \hat n_d
+ \kappa_2 \hat T_2 \cdot \hat T_2$, with $\chi=-\sqrt{7}/2$ for
$N=16$. A sharp transition occurs at a particular value of the ratio 
$-\varepsilon_d /\kappa_2$ at which point the equilibrium value
of $\beta$ drops from a finite value to zero.   

\subsection{Influence of multi-phonon excitations}

Besides the mean-field approximation,
another approximation made in this study
is a truncation in the number of excited phonons
that are included in the basis.
This number is limited to two here.
The previous results indicate that, in the majority of cases,
excited states with different phonon number are decoupled.
It would be of interest, however,
to study the influence of $n$-phonon states with $n>2$.
Some idea of their influence,
without actually extending the phonon basis,
can be obtained by comparing the previous results (which include $n=0,1,2$)
with those where $n=0,1$.
This is done in Figs.~\ref{ppqq}, \ref{t4t4qq} and \ref{edqq}.
The conclusion is that
the influence of two-phonon on one-phonon states is small
except for some extreme parameter values
and in regions where the two cross
({\it e.g.~}for $\kappa_0/\kappa_2\approx-4$ in Fig.~\ref{ppqq}).

\subsection{Simulation of triaxiality in mean field theories}
\label{triax}

In the standard IBM,
{\it i.e.~}with hamiltonians up to two-body operators,
a rigid triaxial shape cannot be obtained \cite{Gino80,Diep80,Isac81}.
One way to induce triaxiality is to include
a cubic term \cite{Isac81} into the hamiltonian.
The intrinsic-state formalism supposes the presence of an average field,
the characteristics of which
should depend self-consistently on the hamiltonian.
A simple way to investigate the consequences of triaxiality
is to relax the self-consistency requirement
and to impose by hand a triaxial mean field through a value $\gamma\neq0^o$.
It must be emphasized that in mean-field theories
it is allowed to perturb the average field
coming from the minimization of the energy.

In Fig.~\ref{var-gam} a calculation of this type
is illustrated for a simple hamiltonian varying $\gamma$.
The value of $\beta$ is obtained minimizing the energy for a fixed $\gamma$.
For $\gamma>15^o$ the lowest band loses the character of ground state,
{\it i.e.~}this band is no longer a condensate of $\Gamma_c$ bosons,
and the interpretation of the levels becomes more complicated.
A intriguing feature of this figure
is that the excitation energies of the two-phonon states
are approximately constant
whereas the excitation energies of one-phonon states
decrease smoothly with $\gamma$.
In other words a sizeable anharmonicity occurs for $\gamma\neq0$.
For example, for $\gamma\sim 6^o$
the anharmonicity $\varsigma$ (\ref{anhar}) is around $0.5$.
This degree of anharmonicity is observed in $^{166}$Er \cite{Garr97,Fahl96}
and $^{168}$Er \cite{Cast80,Born91}.

This result points towards a link between anharmonicity and triaxiality.
For a better understanding of this connection, however,
a fully self-consistent calculation should be carried out
which includes three-body interactions that induce triaxiality.

\section{Electromagnetic transitions}
\label{el-trans}

To calculate electromagnetic transitions in the intrinsic frame
one must assume that the total wave function
can be separated into intrinsic and rotational (collective) part.
This hypothesis is only strictly true for the $SU(3)$ limit
but it can be considered as a good approximation in well-deformed nuclei.
The majority of deformed nuclei have axially symmetric shapes
with an additional symmetry plane
perpendicular to the nuclear symmetry axis.
In this section we deal with this kind of nuclei.

A general multipole operator, in the laboratory frame, of order $L$
is related with the intrinsic one through
\begin{equation} 
\label{mult-op}
\hat T_{LM} (lab)= \sum_{M'} {\cal D}^L_{M M'} \hat T_{LM'} (int),
\end{equation} 
where ${\cal D}$ are the rotation matrices.
Laboratory and intrinsic states
are also related through the ${\cal D}$ matrices,
\begin{equation}
\label{lab-func}
|J M K,\xi\rangle=\sqrt{{2J+1\over 16\pi^2 (1+\delta_{K0})}}
\left({\cal D}^J_{M K}+(-1)^{J+K}{\cal D}^J_{M -K}{\cal R}_2(\pi)\right) 
|K \xi\rangle,
\end{equation}
where  $\xi$ [cfr.~(\ref{eigfun})]
denotes the intrinsic part of the wave function
and $K$ the projection of the total angular momentum on the symmetry axis.

In our case $|K \xi\rangle $ are the intrinsic states 
(\ref{etas}-\ref{exst}, \ref{g2K0}-\ref{g2K4}) and
reduced matrix elements in the laboratory frame
can be related to the intrinsic matrix elements as follows:
\begin{eqnarray}
\label{ma-elem}
\nonumber
\langle & J &  K,\xi ||T_L (lab) || J' K',\xi'\rangle = 
\sqrt{2 J'+1}  
\left(\langle J' K' \,L K-K'|J K\rangle 
\langle K \xi |\hat T_{L K-K'} (int)|K' \xi'\rangle  \right. \\
& + & \left. (-1)^{J'+K'} \langle J' -K' \, L K+K'|J K\rangle \,
\langle K \xi |\hat T_{LK+K'}|\overline{ K' \xi'}\rangle \right)~
{2 \over 1+\delta_{K0}+\delta_{K'0}+\delta_{KK'}}~. 
\end{eqnarray}
The last factor in r.h.s. is introduced to take into account that our 
intrinsic states contain both components $\pm K$.

The reduced transition probability for a transition
between states belonging to two rotational bands is given by
\begin{equation}
\label{be2}
B(E2,J' K',\xi'\rightarrow J  K,\xi)={1\over2 J' +1}
|\langle J K,\xi ||T_L (lab)|| J' K',\xi'\rangle|^2.
\end{equation}
The knowledge of the intrinsic matrix element
between two rotational bands
hence determines the reduced matrix element (\ref{ma-elem})
from where the reduced transition probability can be deduced.

Of special relevance to Nuclear Physics are E2 transitions
and results given in this section are confined to this type of transition.
The most general one-body E2 operator is defined in Eq.~(\ref{OpeMul})
(with an effective charge equal to $1$).
Matrix elements between the ground state and one-phonon states
are given in Ref.~\cite{Alo92}.
Likewise, matrix elements between one- and two-phonon can be derived.
The non-vanishing ones are given by (we keep the $\gamma$--dependence
for generality)
\begin{equation}
\langle \beta |\hat T_{20}|\beta ^2\rangle = 
\sqrt {2(N-1)}~ {(1-\beta^2)\cos \gamma  -\sqrt {2 \over 7} 
\beta \chi \cos 2\gamma \over {1+\beta ^2}},
\end{equation}

\begin{equation}
\langle \beta |\hat T_{2\pm 2}|\beta ^2\rangle =
\sqrt {2(N-1)}~{ {1 \over \sqrt{2}}(1-\beta^2) \sin\gamma+
{1\over\sqrt{7}}\beta\chi\sin 2\gamma
\over {1+\beta^2}},
\end{equation}

\begin{equation}
\langle \beta |\hat T_{20}| \beta \gamma_+ \rangle =
\sqrt{N-1}~ {\displaystyle {-\sin\gamma +  \sqrt{2\over 7} 
\beta \chi \sin 2\gamma 
\over \sqrt{1+\beta^2}}},
\end{equation}

\begin{equation}
\langle \beta |\hat T_{2\pm2}| \beta \gamma_+ \rangle =
\sqrt{N-1}~ {{{\displaystyle 1 \over \sqrt{2}} \cos \gamma + 
{\displaystyle 1 \over \sqrt 7 } \beta  \chi  \cos 2\gamma} 
\over{\sqrt{1+\beta^2}}},
\end{equation}

\begin{equation}
\langle \gamma_+|\hat T_{20}|\beta \gamma_+\rangle =
{1\over\sqrt{2}}\langle \beta |\hat T_{20}|\beta ^2\rangle,  
\end{equation}

\begin{equation}
\langle \gamma_+|\hat T_{2\pm 2}|\beta \gamma_+\rangle =
{1\over\sqrt{2}}\langle \beta |\hat T_{2\pm 2}|\beta ^2\rangle,  
\end{equation}

\begin{equation}
\langle \gamma_+|\hat T_{20}| \gamma^2_{K=0} \rangle =
\langle \gamma_+|\hat T_{20}| \gamma^2_{K=4} \rangle =
\langle \beta |\hat T_{20}| \beta \gamma_+ \rangle ,
\end{equation}

\begin{equation}
\langle \gamma_+|\hat T_{2\pm2}| \gamma^2_{K=0} \rangle =
\langle \gamma_+|\hat T_{2\pm2}| \gamma^2_{K=4} \rangle =
\langle \beta |\hat T_{2\pm2}| \beta \gamma_+ \rangle,
\end{equation}

\begin{equation}
\langle \beta^2 |\hat T_{20} |\beta^2\rangle = 
 (N-2) {{2 \beta \cos \gamma_-\sqrt {2 \over 7} \beta^2 \chi
\cos 2\gamma} \over {1+\beta^2}}+
2 {{-2 \beta \cos \gamma -\sqrt{2 \over 7} 
\chi \cos 2\gamma}\over {1+\beta^2}},
\end{equation}

\begin{equation}
\langle \beta^2 |\hat T_{2\pm 2} |\beta^2\rangle =
(N-2){\displaystyle{\sqrt{2}\beta\sin\gamma+{1\over\sqrt{7}}\beta^2 \chi 
\sin 2\gamma \over{1+\beta^2}}}\\
+2{\displaystyle
{-\sqrt{2}\beta \sin\gamma +{1\over\sqrt{7}}\chi\sin 2\gamma \over{1+\beta^2}}},
\end{equation}

\begin{equation}
\langle \beta^2 |\hat T_{20}|\beta \gamma_+\rangle=
{\sqrt{2}\beta \sin\gamma +{2\over\sqrt{7}}\chi\sin 2\gamma     
\over{\sqrt {1+\beta^2}}},
\end{equation}

\begin{equation}
\langle \beta^2 |\hat T_{2\pm2}|\beta \gamma_+\rangle= 
{{- \beta \cos \gamma 
+\sqrt {2 \over 7} \chi \cos 2\gamma} \over 
{\sqrt {1+\beta^2}}},
\end{equation}

\begin{eqnarray}
\langle \beta \gamma_+ |\hat T_{20}|\beta \gamma_+\rangle=
(N-2) {\displaystyle {{2 \beta \cos \gamma 
-\sqrt {2 \over 7} \beta^2 \chi \cos 2\gamma} \over 
{1+\beta^2}}}
& + & {\displaystyle{{-2 \beta \cos \gamma - \sqrt{2 \over 7} \chi 
\cos 2\gamma}\over {1+\beta^2}}} \\ \nonumber
& + & \sqrt {2 \over 7} \chi  \cos 2\gamma,
\end{eqnarray}

\begin{eqnarray}
\langle \beta \gamma_+ |\hat T_{2\pm 2}|\beta \gamma_+\rangle=
(N-2){\displaystyle{\sqrt{2} \beta\sin\gamma +{1\over\sqrt{7}}
\beta^2 \chi \sin 2\gamma
\over{1+\beta^2}}}
& + &{\displaystyle
{-\sqrt{2}\beta \sin\gamma +{1\over\sqrt{7}}\chi\sin 2\gamma 
\over{1+\beta^2}}} \\ \nonumber
& - & {\displaystyle{\chi\sin 2\gamma\over \sqrt{7}}},
\end{eqnarray}

\begin{equation}
\langle \beta \gamma_+ |\hat T_{20}|\gamma^2_{K=0} \rangle =
\langle \beta \gamma_+ |\hat T_{20}|\gamma^2_{K=4} \rangle =
 {1\over \sqrt{2}} \langle \beta^2 |\hat T_{20}|\beta \gamma_+\rangle,
\end{equation}

\begin{equation}
\langle \beta \gamma_+ |\hat T_{2\pm 2}|\gamma^2_{K=0} \rangle =
\langle \beta \gamma_+ |\hat T_{2\pm 2}|\gamma^2_{K=4} \rangle =
 {1\over \sqrt{2}} \langle \beta^2 |\hat T_{2\pm2}|\beta \gamma_+\rangle, 
\end{equation}

\begin{equation}
\langle \gamma^2_{K=0} |\hat T_{20}|\gamma^2_{K=0} \rangle=
\langle \gamma^2_{K=4} |\hat T_{20}|\gamma^2_{K=4} \rangle=
(N-2) {\displaystyle {{2 \beta \cos \gamma 
-\sqrt {2 \over 7} \beta^2  \chi \cos 2\gamma} \over 
{1+\beta^2}}}+ \sqrt {2 \over 7} \chi (\cos 2\gamma +1),  
\end{equation}

\begin{equation}
\langle \gamma^2_{K=0} |\hat T_{2\pm2}|\gamma^2_{K=0} \rangle=
\langle \gamma^2_{K=4} |\hat T_{2\pm2}|\gamma^2_{K=4} \rangle=
(N-2) {\displaystyle {{\sqrt{2} \beta \sin \gamma 
+\sqrt {1 \over 7} \beta^2 \chi \sin 2\gamma} \over 
{1+\beta^2}}}- \sqrt {1 \over 7} \chi \sin 2\gamma ~. 
\end{equation}

In order to see the general behaviour of the calculated transitions,
in Fig.~\ref{transit} we present a calculation for the $B(E2)$'s 
as a function of $\chi$ for the hamiltonian
$H = \kappa_2 \left(\hat T_2 \cdot \hat T_2 - \hat P^{\dagger} \hat P \right)$.
In (a) the calculation is done directly in the lab system while in
(b) the intrinsic wave functions and Eqs. (\ref{ma-elem}-\ref{be2}) are used. 
The main characteristics of the calculated $B(E2)$'s are
similar to those obtained in the lab in Ref.~\cite{Cast88} although in that 
case only ground state and one phonon excitations are included.

\section{Conclusions}
\label{conclu}

In this paper a formalism is presented
for performing calculations with up to two phonon-excitations
in the intrinsic frame.
This formalism yields good agreement with exact results
and allows an easy interpretation of them.
The calculations can be easily extended
to include higher multi-phonon excitations
and/or higher-order interactions.

From these results it is deduced
that for a large number of bosons $N$
the IBM is a harmonic model even if $n$-body interactions are included.
A sizeable degree of anharmonicity can be obtained
only for a finite number of bosons
combined with a three-body term in the hamiltonian.
This result points towards a link
between triaxiality and anharmonicity in the IBM.

\section{Acknowledgments}
We acknowledge Prof. F. Iachello for a careful reading of the manuscript.
This work has been supported in part by the Spanish DGICYT under contract
No. PB95--0533, by the European Commission under contract CI1*-CT94-0072
and by IN2P3 (France)-CICYT (Spain) and INFN (Italy)-CICYT (Spain)
agreements.

\begin{figure}[]
\caption{Energy levels as a function of the number of bosons $N$ 
for the $SU(3)$ limit. The hamiltonian $H=k_2
\hat T_2\cdot\hat T_2$
is used. Full lines correspond to the intrinsic-frame 
calculations and dashed lines to the laboratory-frame calculations.}
\label{fig-su3}
\end{figure}
   
\begin{figure}[]
\caption{Energy levels as a function of the number of bosons $N$ 
for the $O(6)$ limit. The hamiltonian $H=k_0
\hat P^\dagger \hat P$
is used. Full lines correspond to the intrinsic-frame 
calculations and dashed lines to the laboratory-frame calculations.}
\label{fig-o6}
\end{figure}

\begin{figure}[]
\caption{Energy levels and composition of states as a function
of the ratio $-\kappa_0/\kappa_2$ for the hamiltonian 
$H=\kappa_0\hat P^{\dagger} \hat P + \kappa_2 \hat T_2 \cdot \hat T_2$, 
with $\chi=-\sqrt{7}/2$ and $N=16$
for up to (a) two phonons and (b) one phonon.}
\label{ppqq}
\end{figure}

\begin{figure}[]
\caption{Energy levels and composition of the states 
as  a function of the ratio $-\kappa_4/\kappa_2$ for the
hamiltonian 
$H=\kappa_2 \hat T_2 \cdot \hat T_2 +\kappa_4 \hat T_4 \cdot \hat T_4$, 
with $\chi=-\sqrt{7}/2$ and $N=16$
for up to (a) two phonons and (b) one phonon.}
\label{t4t4qq}
\end{figure}

\begin{figure}[]
\caption{Energy levels and composition of the states 
as  a function of the ratio $-\varepsilon_d/\kappa_2$ for the
hamiltonian $H=\varepsilon_d \hat n_d +\kappa_2 \hat T_2 \cdot \hat T_2$, with
$\chi=-\sqrt{7}/2$ and $N=16$
for up to (a) two phonons and (b) one phonon.}
\label{edqq}
\end{figure}

\begin{figure}[]
\caption{Excitation energies and composition of the states 
as a function of $\gamma$ 
in a $\gamma$-perturbed mean-field
calculation for the hamiltonian
$H=\kappa_2(\hat T_2 \cdot \hat T_2-\hat P^\dagger P)$, with
$\chi=-\sqrt{7}/2$ and $N=16$.}
\label{var-gam}
\end{figure}

\begin{figure}[]
\caption{$B(E2)$ values for selected transitions between ground, one- 
and two-phonon excited states for a hamiltonian 
$H = \kappa_2 \left(\hat T_2 \cdot \hat T_2 - \hat P^{\dagger} \hat P \right)$ 
as a function of $\chi$ for a system with $N=16$.
In (a) the calculation is done directly in the lab system and in (b)
the intrinsic wave functions and Eqs. (\ref{ma-elem}-\ref{be2}) are used.}
\label{transit}
\end{figure}

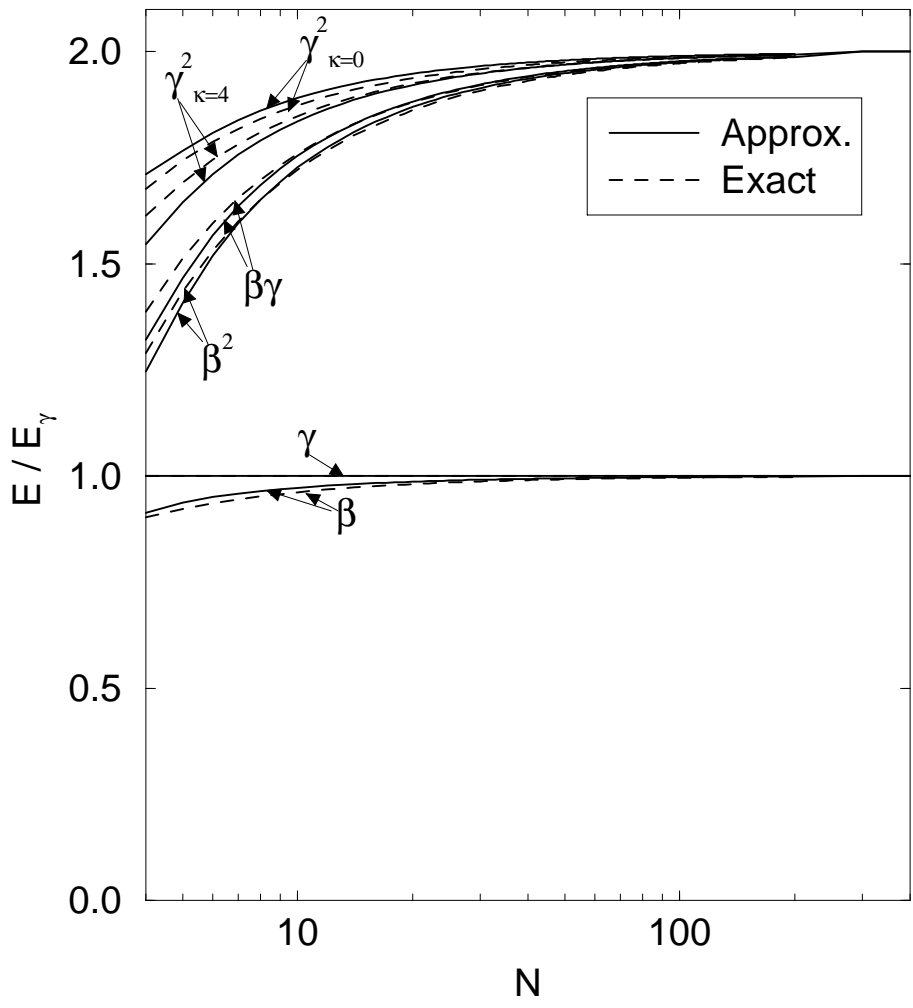
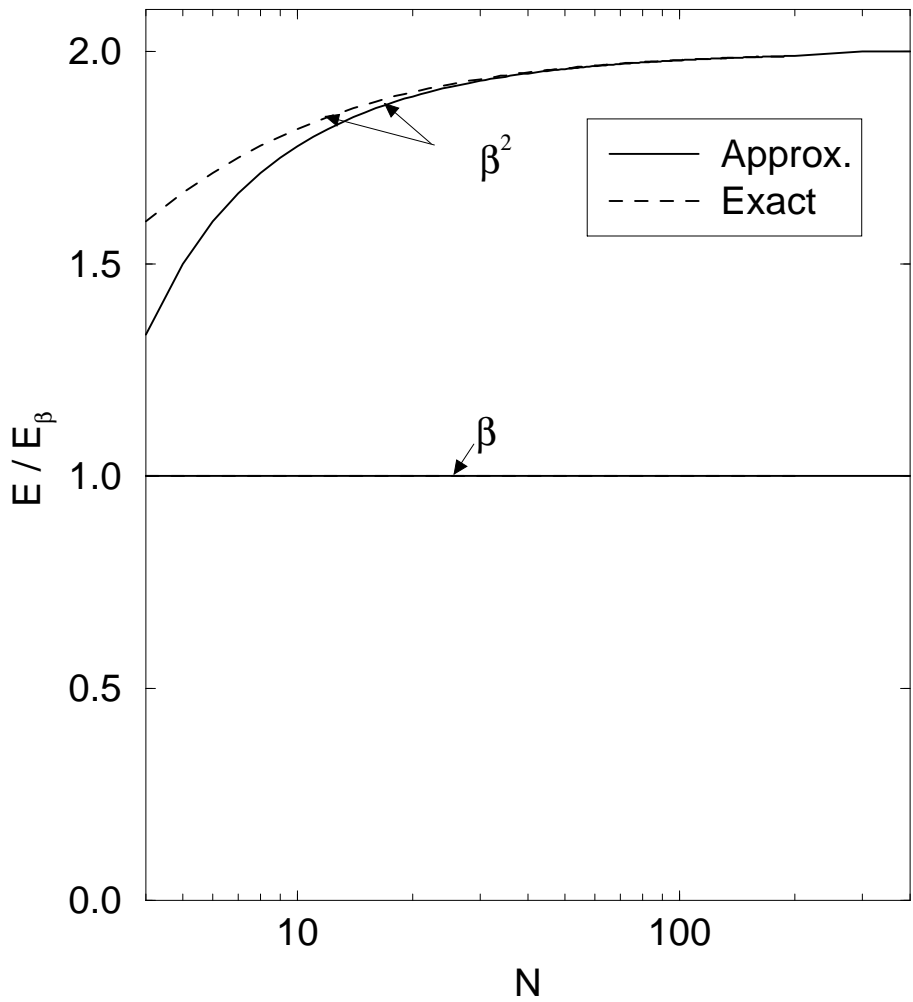
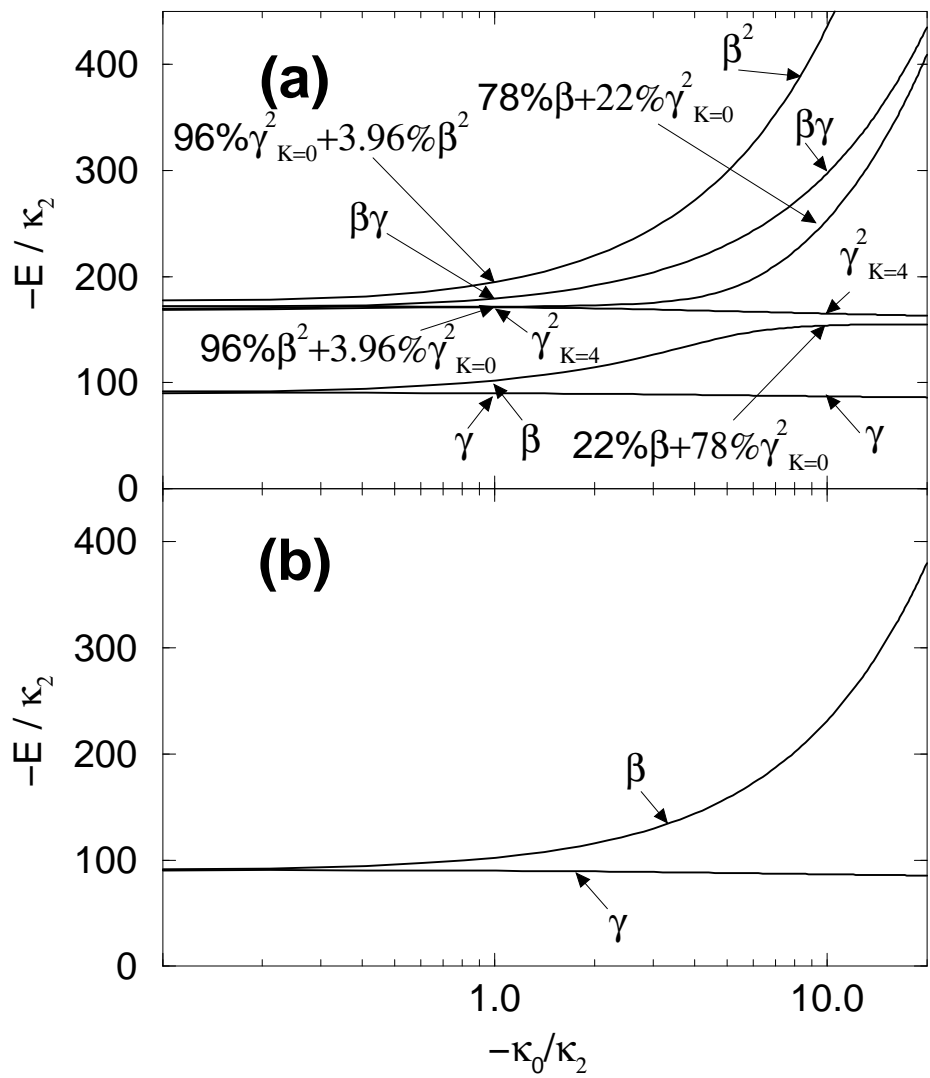
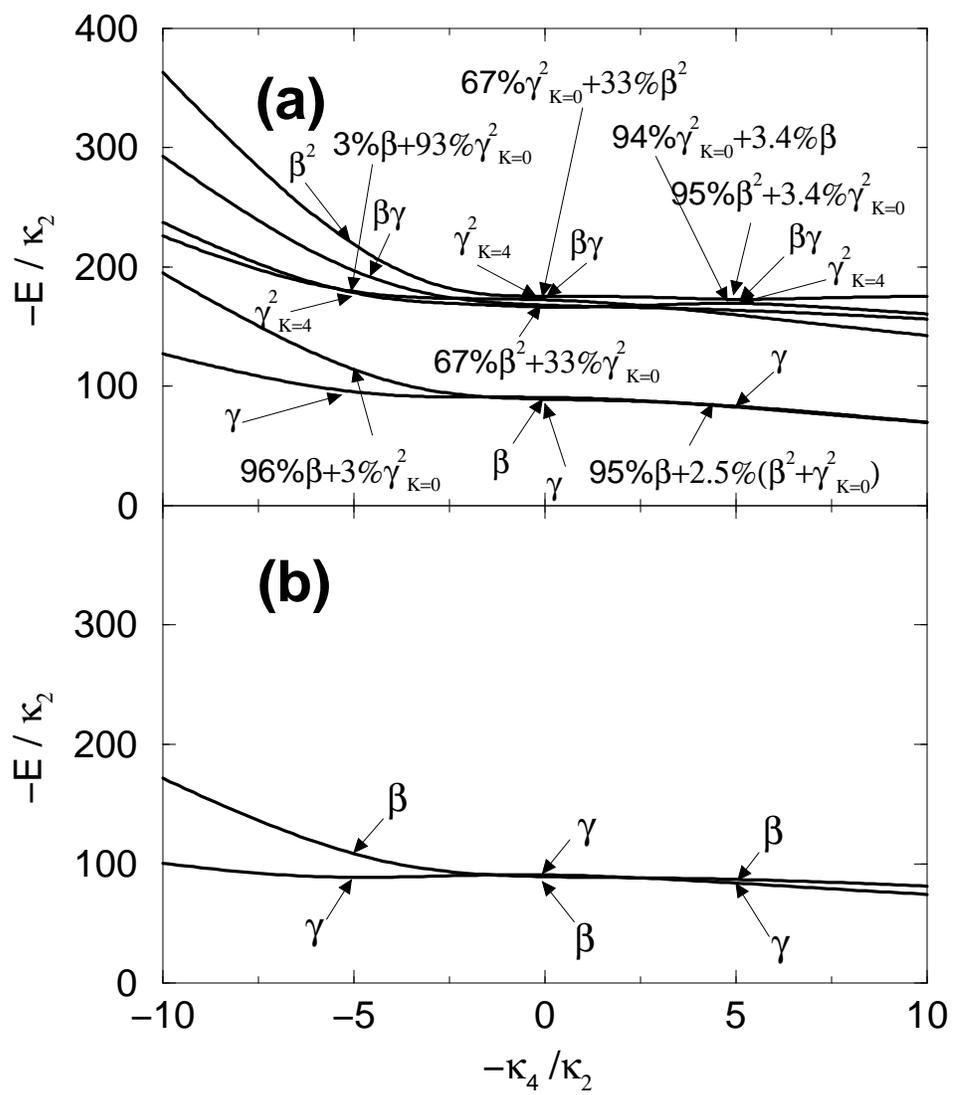
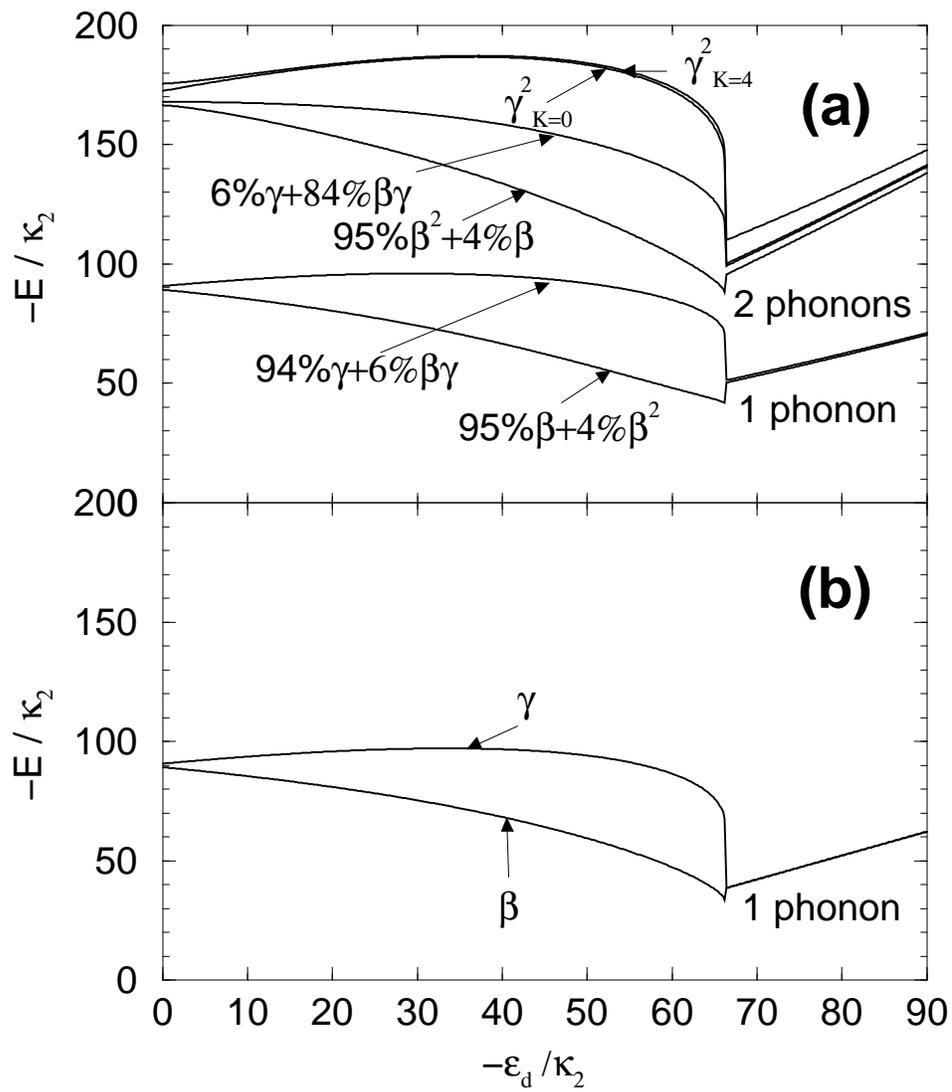
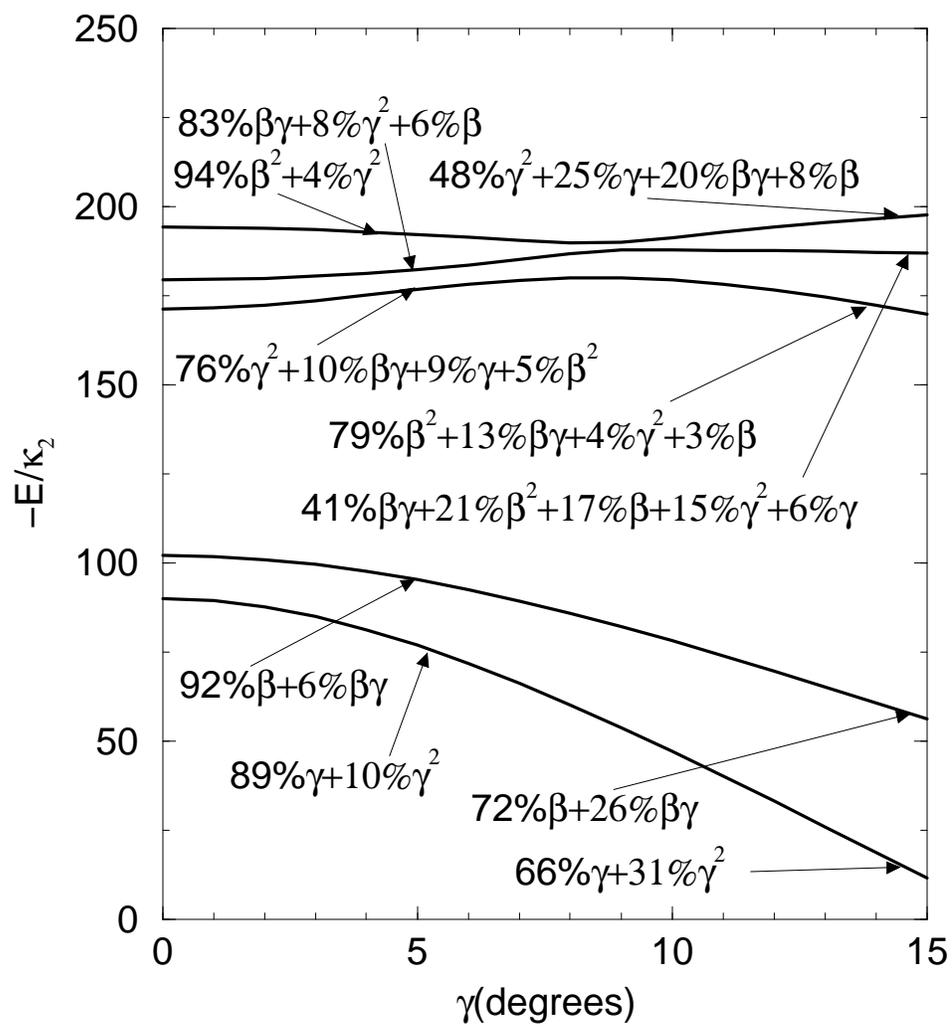
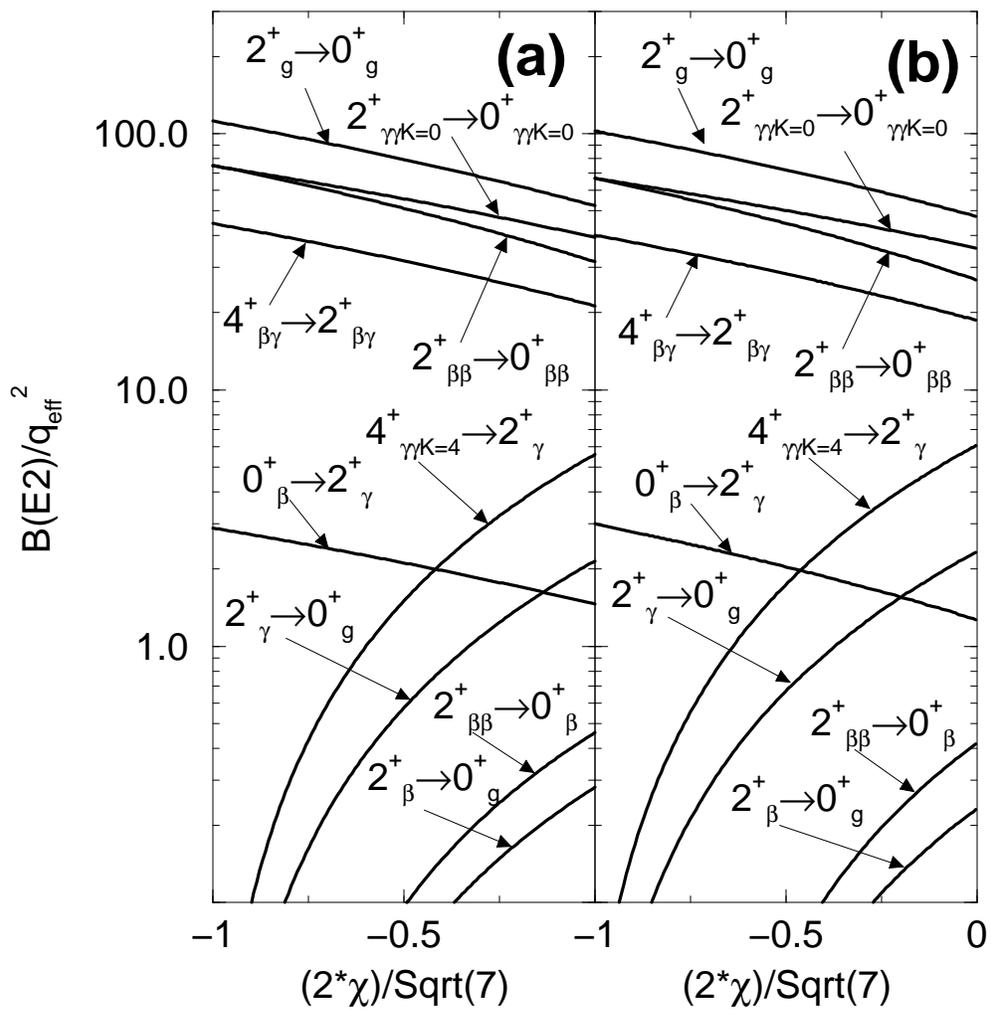
\end{document}